\documentclass[usenatbib,usegraphicx]{mn2e}

\usepackage{multirow}

\title[A new star-forming region in CMa]{A new star-forming region in Canis Major}
\author[T. Yu. Magakian, T.A. Movsessian and J. Bally]{T. Yu. Magakian$^{1}$\thanks{E-mail:
tigmag@sci.am} T.A. Movsessian$^{1}$ and J. Bally$^{2}$\\
$^{1}$Byurakan Observatory, Aragatsotn reg., 0213, Armenia\\
$^{2}$Center for Astrophysics and Space Astronomy, University of Colorado, 391 UCB, Boulder, CO 80309-0001, USA}
\begin{document}

\date{Accepted ... Received ...; in original form ...}

\pagerange{\pageref{firstpage}--\pageref{lastpage}} \pubyear{2016}

\maketitle

\label{firstpage}

\begin{abstract}
A new southern star-formation region, located at an estimated distance of $\sim$1.5 kpc 
in the Lynds 1664 dark cloud in Canis Major,  is described.    
Lynds 1664  contains several compact star clusters, small stellar groups, and young 
stars associated with  reflection nebulae.   Narrow-band H$\alpha$ and [SII] images 
obtained with 4-m CTIO telescope reveal more than 20 new Herbig-Haro 
objects associated with several protostellar outflows. 
\end{abstract}


\section{Introduction}

A group of the several faint nebulous objects were discovered around RA = 7h 24m and Dec = $-$24d 30m during a search for cometary nebulae and Herbig-Haro (HH) objects  on the Palomar Observatory Sky Survey (POSS)  images \citep{GM1,GM2,GM3}.     These nebulae are associated with the Lynds~1664 (L1664) dark cloud far from other well-known star forming complexes in Canis Major such as  CMa OB1 and CMa R1.  At the time of their discovery, the large negative declination prevented the acquisition of  better  data.    Early surveys of this field identified several reflection nebulae that were listed in \citet{BBWo}.   However, in the subsequent decades, no additional data became available and these objects remained unstudied.    In this paper, we present  deep narrow-band H$\alpha$ and [SII]  images of the L1664 cloud and report  the discovery of several new HH objects tracing outflows from  small clusters of young stellar objects (YSOs) detected on WISE satellite infrared images.    

\section[]{Observations}

The images presented here were obtained on the nights of 13 May 2004 
using the NOAO Mosaic II Camera CCD camera at the f/3.1 prime focus 
of the 4 meter Blanco telescope at the Cerro Tololo Interamerican 
Observatory (CTIO) near La Serana, Chile.   Mosaic II camera is a 
8192$\times$8192 pixel array (consisting of eight 
2048$\times$4096 pixel CCD chips) with a pixel scale of 
0.26$''$ pixel$^{-1}$ and  a field of view 35.4$'$ on a side.  

Narrow-band filters centered on 
6569\AA\ and 6730\AA\   with a FWHM bandwidth 
of 80\AA\  were used to obtain  H$\alpha$ and [SII] images.  
A Sloan Digital Sky Survey (SDSS) i'  filter centered on 
7732\AA\ with a FWHM of 1548\AA\  was used for continuum imaging.
A set of five dithered 600 second exposures were obtained in 
H$\alpha$ and [SII] using the standard MOSDITHER pattern 
to eliminate cosmic rays and the gaps between the
individual chips in Mosaic.  A dithered set of five 180 second 
exposures were obtained  in the broad-band SDSS i-band filter to
discriminate between H$\alpha$, [SII], and continuum emission. 

Images were reduced in the standard manner using IRAF.  Following
bias subtraction, cosmic ray removal, and flat fielding using dome flats, 
images were combined using the MSCRED package in IRAF.

\begin{figure*}
\includegraphics{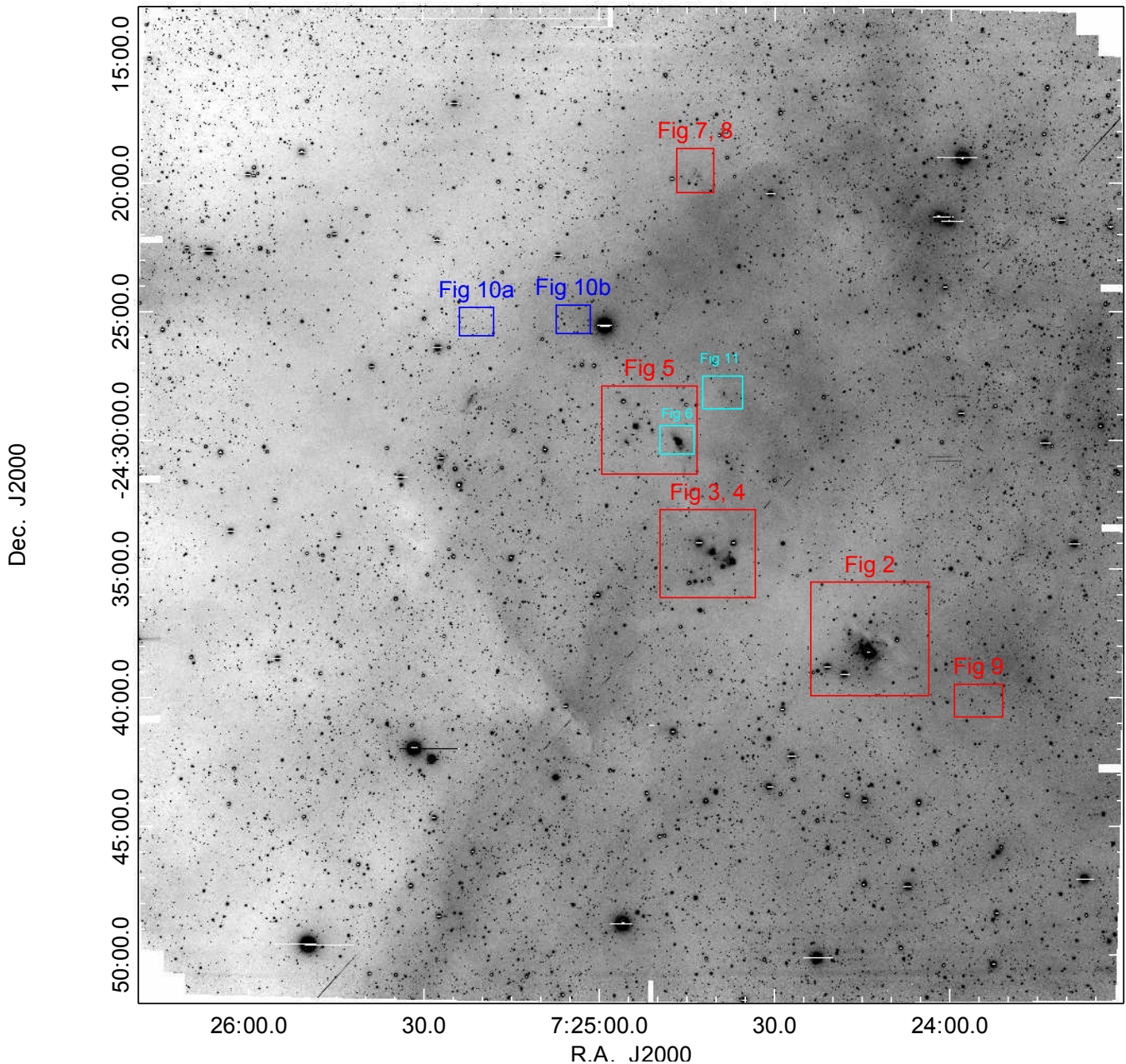}
\caption{Whole observed field in H$\alpha$ filter. The areas, which are discussed in detail below, are shown by rectangles.} 
\label{field}
\end{figure*} 

\section[]{Results}

\begin{table*}
\label{HHO_table}
 \centering
 \begin{minipage}{140mm}
  \caption{The coordinates of HH objects, HH flows and other emission objects in the field.}
  \begin{tabular}{llll@{}}
  \hline
   Name     &  RA(2000)    & Decl.(2000) & Notes  \\
 \hline
  HH 1108 & 7 23 55.1 & $-$24 39 57   & Streak with condensations \\
  HH 1109 & 7 24 11.9 & $-$24 38 33   & To the SW from BBWo 22C \\
  HH 1110 & 7 24 18.4 & $-$24 36 42   & To the NE from BBWo 22C \\
  HH 1112 & 7 24 41.5 & $-$24 34 15   & To the NE from GM 2-20 \\
  HH 1111A & 7 24 43.6 & $-$24 19 32   & \multirow{4}{*}{$\Biggr\}$A separate flow from embedded source}\\
  HH 1111B & 7 24 43.2 & $-$24 19 46   & \\
  HH 1111C & 7 24 43.0 & $-$24 19 49   & \\
  HH 1111D & 7 24 42.6 & $-$24 19 45   & \\
  HH 1113 & 7 24 45.2 & $-$24 33 40   & To the NE from GM 2-20 \\
  HH 1114 & 7 24 45.4 & $-$24 31 13   & Star-like HH knot, connected with reflection nebula\\

  HH 1120A & 7 24 46.2 & $-$24 33 05   &  Faint spot \\
  HH 1120B & 7 24 48.0 & $-$24 33 00   & \multirow{2}{*}{$\Bigr\}$Two faint knots} \\
  HH 1120C & 7 24 48.3 & $-$24 32 59   &  \\
  HH 1116 & 7 24 50.2 & $-$24 29 23   & Two faint knots \\
  HH 1115 & 7 24 51.5 & $-$24 29 23   &  HH jet from the star inside GM 2-21 nebula.    \\
  HH 1117 & 7 24 55.3 & $-$24 28 39   &  Small knot     \\   
  HH 1118A & 7 24 56.6 & $-$24 28 10   &  \multirow{3}{*}{$\biggr\}$Several bright knots in common nebula}    \\
  HH 1118B & 7 24 56.8 & $-$24 28 11   &   \\
  HH 1118C & 7 24 57.0 & $-$24 28 11   &  \\
HH 1119A & 7 24 58.0 & $-$24 31 12 & \multirow{2}{*}{$\Bigr\}$ A dumbbell-like pair of HH knots} \\ 
HH 1119B & 7 24 58.7 & $-$24 31 10\\ 

\hline
  Em.Obj. 1 & 7 25 04.4 & $-$24 25 19 \\ 
  Em.Obj. 2 & 7 25 20.9 & $-$24 24 53 \\
 
\hline
\end{tabular}
\end{minipage}
\end{table*}

\subsection{HH objects and flows}

Herbig-Haro objects are shocks powered by protostellar outflows.  They can be identified as compact H$\alpha $ and  6716/6731\AA\ [SII]  emission line sources without continuum emission in the SDSS i'  band images.    HH objects  tend to have [SII]  / H$\alpha\ $ surface-brightness ratios larger than  $\sim$0.5 while photo-ionized clumps have ratios smaller than 0.2.  Thus, the intensity ratios of these emission lines, along with the presence or absence of i' band emission, can be used to discriminate between HH objects, photo-ionized gas, and reflection nebulae.

The area under investigation contains many small dark clouds in which several new groups of the HH objects were identified.    However,  outflow activity traced by  HH objects is confined to a small portion of the L~1664 cloud.    Figure \ref{field} shows a wide-field  H$\alpha\ $ image of the region. 
Most detected HH, listed in Table 1  in the order of their right ascension,  are concentrated along a northeast-southwest  line connecting three groups of small reflection nebulae around BBWo~22C, GM~1-8,  and GM~3-7 nebulae.     The new HH objects and their environments  are described below.  

\subsubsection{The BBWo~22C group}

\begin{figure}
\includegraphics[width=80mm]{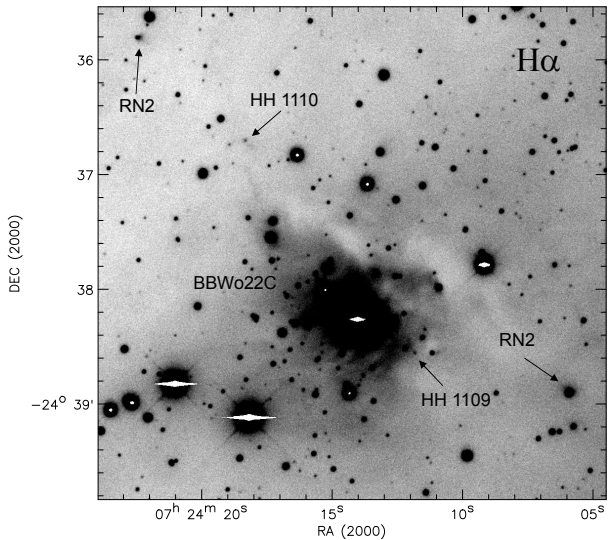}
\caption{An image of the area around BBWo~22C in H$\alpha$. HH~1109, HH~1110 and other objects are marked.} 
\label{BBWo_22C}
\end{figure}

BBWo~22C (Figure \ref{BBWo_22C}) is a reflection nebula, located near  the center of the L1664 dark cloud,  that contains  a small open cluster consisting of two dozen stars.     The cluster was found by  \citet{ivanov} (numbered as CC~06) and by \citet{DBS} (numbered as DBS~11) in the 2MASS survey.  The cluster and reflection nebula are associated with the infrared source IRAS~07221-2531.   The cluster is rimmed by several opaque cloudlets along its northern and western periphery.  None of the stars seen at visual wavelengths appear to be particularly reddened and 2MASS does not reveal any objects seen only at infrared wavelengths.    Two compact  HH objects were found near this cluster (see Figure~\ref{BBWo_22C}). HH~1109 is brighter in H$\alpha$ than [SII] and exhibits a short tail extending towards the northeast which may indicate a bow-shock propagating towards the southwest.   HH~1110 is an unresolved emission-line knot  brighter in H$\alpha$ than [SII]  located northeast of the cluster.    The source(s) of these HH objects could either be cluster members or highly obscured young stellar objects embedded in the adjacent dark clouds.      Proper motion measurements are needed to indicated the approximate locations of the driving sources.  

This portion of L1664 contains two nebulous stars,  RN~1 and RN~2,  shown  in Figure~\ref{BBWo_22C};   their coordinates are listed in Table 2 with other reflection nebulae in this cloud.    RN1 exhibits a fan of illumination toward the north;  RN2 exhibits a fan of light extending from a dim star towards the west.     Although both objects are visible in the WISE 3.6 and 4.5 $\mu$m  images, neither is particularly prominent at these wavelengths.    Comparison of the fluxes in the i'-band with the infrared images obtained by the Wide-field Infrared Survey Explorer (WISE) shows that RN2 is highly reddened but RN1 is not. 

\subsubsection{The GM~1-8 group}

\begin{figure}
\includegraphics[width=80mm]{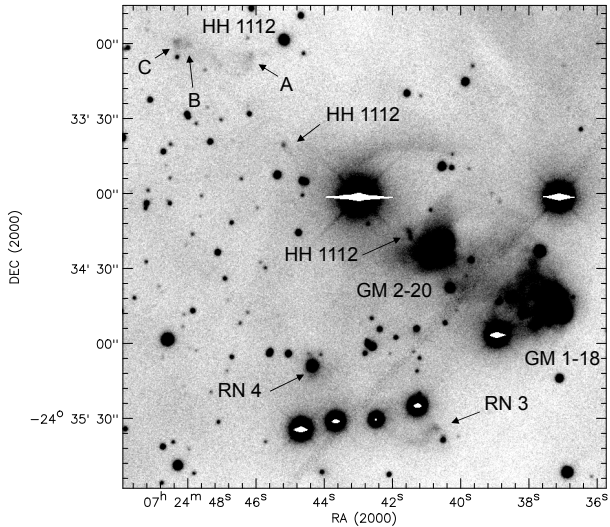}
\caption{GM~1-8 and GM~2-20 area in H$\alpha$. HH~1112, 1113, 1120 and reflection nebulae are marked.}
\label{GM2-20}
\end{figure}

\begin{figure}
\includegraphics[width=80mm]{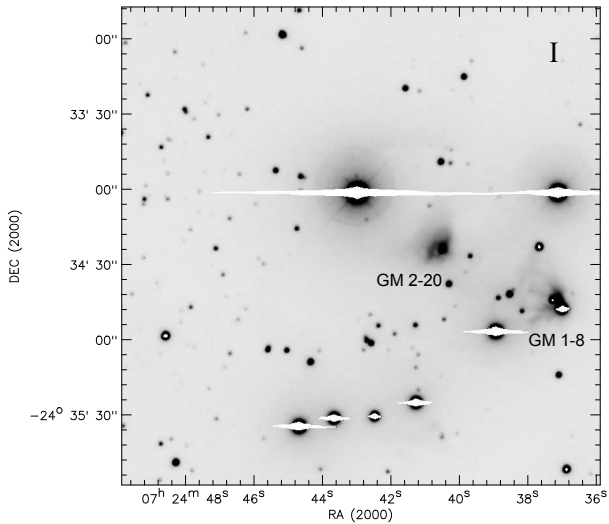}
\caption{The same area as in Figure \ref{GM2-20} in i'  filter and with lower contrast, showing the inner structure of GM~1-8 and GM~2-20 nebulae, as well as the duplicity of the star in GM~1-8.}
\label{GM2-20i}
\end{figure}

The GM~1-8 group (Figs. \ref{GM2-20} and \ref{GM2-20i}) contains  two prominent northeast-facing cometary nebulae: GM~1-8 and GM~2-20 (BBWo~22E).   Several nebulous streaks point radially away from the star located at the head of the GM~1-8 nebula.  The star at  the apex  of GM~1-8 was included in the Catalog of Galactic OB Stars \citep{reed1} as ALS~19644. It coincides with IRAS~07225-2428 \citep{glmp} (catalogued as GLMP~181) and is likely to be a moderate mass pre-main-sequence (PMS) object.     Spectroscopy shows that  H$\alpha$ along with the forbidden emission lines  are in emission  \citep{pds1,pds2},  confirming that it is a moderate-mass HAeBe star.    It is very bright on the WISE images  in all  bands and its flux increases towards longer wavelengths.  Thus this star must be surrounded by a dusty envelope.     
  Another, fainter  star is  embedded in GM~1-8 approximately 4\arcsec\ to the northeast of the HAeBe star.  However, no data constraining the nature of this object exists and it may be unrelated to star formation activity in L~1664 (it is not conspicuous at infrared wavelengths).   

The nebula  GM~2-20 also has a cometary morphology  with the star in the tip of a  red cone with same orientation as GM~1-8.  The nebula contains a small nebulous arc emerging from the star. Compared to GM~1-8, its the star is not  bright in the WISE images especially at longer wavelengths. The SIMBAD  coordinates for GM~2-20 and BBWo~22E are in error;  these two designations refer to the  same object.  \citet{soares1} proposed that nearly all stars surrounding the GM~2-20 and GM~1-8 nebulae form a small cluster based on 2MASS images; the WISE  images  confirm this suggestion.

HH~1112 (Table 1), an oblong emission knot  brighter in  H$\alpha$ than [SII],  is located near the northeast edge of the GM~2-20 reflection nebula.  Another  H$\alpha$-dominated HH knot  located further to northeast is designated HH~1113.     HH~1120 is en extended object consisting of three diffuse knots (marked in Figure \ref{GM2-20} as A, B and C) connected by a faint filament.   Knots B and C are separated by only 4" to 5".     HH~1112,  HH~1113,  HH~1120 and the central star in GM~2-20 lie along a straight line.   Thus,  this star may be the source of an outflow driving these HH objects.     HH~1120 resembles a fragmented bow shock at the end of a 2 arcminute-long outflow from from the star embedded in GM~2-20.  

Two additional reflection nebulae, RN~3 and RN~4 located approximately 1 arc-minute south of GM2-20,  are marked in Figure \ref{GM2-20}.     RN~3 is a faint, diffuse filamentary structure opening towards the northeast and possibly connected to a visually dim star that is bright in the K-band and WISE images.     RN~4 is a star surrounded by dim nebulosity consisting of a mixture of  emission and reflected light that is not conspicuous in IR images.   A small knot to north-northeast of this star that is brighter in H$\alpha$ than [SII]  may be another very faint HH object at $7^{\rmn{h}} 24^{\rmn{m}} 44\fs5 -24\degr\ 35\arcmin\ 04\arcsec\  ~(2000.0)$.
 
\subsubsection{GM~3-7 group}

GM~3-7 (BBWo~22F) is a compact nebula  surrounding several stars (Figure  \ref{GM3-7})  located further to the northeast (SIMBAD incorrectly listed it as  identical to
GM~2-21; see below).  No HH objects were detected in this object whose inner structure is shown in Figure  \ref{GM3-7large}.     At visual wavelengths,  it includes two stars of nearly equal brightness (marked 1 and 2),  as well as a fainter object  (star 3).    Star 2 is nebulous and bright in the K$_{s}$ image;   star 4 is also bright in the IR.

\begin{figure}
\includegraphics[width=80mm]{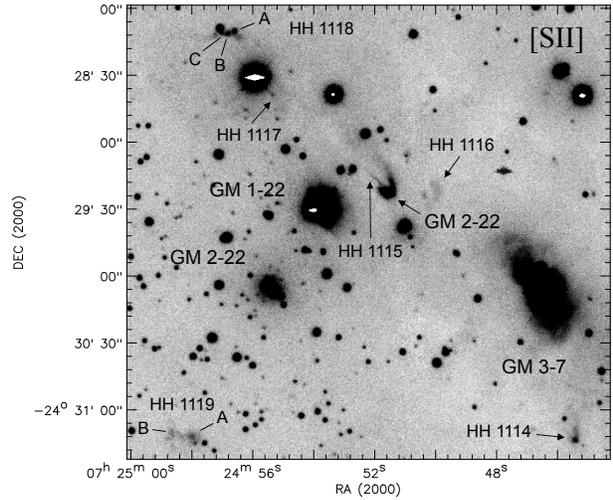}
\caption{A [SII] image of the area surrounding GM~1-22 and GM~3-7. The HH objects, HH jets and reflection nebulae are labelled. Also note that GM~2-22 is a galaxy ZOAG~G238.44-04.08.}
\label{GM3-7}
\end{figure}

\begin{figure}
\includegraphics[width=84mm]{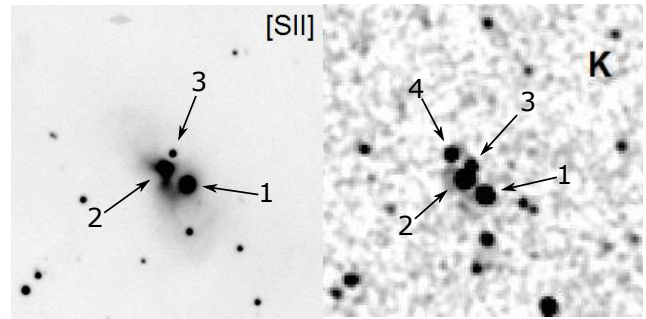}
\caption{Enlarged images of GM~3-7, showing its inner structure in [SII] (left panel) and in K$_{s}$ from 2MASS (right panel). Several embedded stars are labelled. The approximate size of each image is 105$''$$\times$105$''$.}
\label{GM3-7large}
\end{figure} 

GM~1-22 (BBWo~22H) is a bright star on the edge of a dark cloud that illuminates compact fan-shaped reflection nebula with high surface brightness.     There are no  detected HH objects in this region. 

During our early searches on POSS images, two  small red nebulae were found near southeast and northwest sides of GM~1-22 \citep{GM2} and thought to be candidate HH objects and listed as GM~2-21 and GM~2-22.   GM~2-22 (which is incorrectly labeled in SIMBAD as the same object as GM~1-22) is actually a galaxy which is listed as a zone of  avoidance  object  ZOAG~G238.44-04.08.   GM 2-21 is a classical cone-shaped cometary nebula with a red star in the tip of the cone.  It is associated with IRAS~07227-2423 located only in 11\arcsec\  from this star. Its high brightness in the WISE images supports this identification.

Several HH objects were found in this field (Figure~\ref{GM3-7}).  HH~1114 is a nearly star-like emission knot  dominated by  [SII] emission  but also visible in H$\alpha$.     It is accompanied by nebulous wisps of reflection nebulosity which are bright  in the i'  image.     HH~1114 resembles a cometary nebula oriented roughly southwest-northeast with an emission-line core.   The nebulosity is faintly detected in the J and H 2MASS images but in K$_{s}$ a star-like object coincides with the emission-line knot. This central star becomes more prominent at longer wavelengths in the WISE images.      Thus HH~1114 is likely to be powered by a star heavily embedded in  circumstellar dust.      The existence of reflection nebula shows that some starlight escapes the dust envelope.

HH~1115, similar to  HH~1114, is well-defined  in [SII] and exhibits a collimated jet extending along the axis of the GM~2-21 nebula from the southwest to the northeast. 
        
HH~1116 is a faint spot located west of GM~2-21 visible mainly in [SII].   HH~1117 is  a compact knot brighter in H$\alpha$.    HH~1118 is similar in appearance in both  [SII] and in H$\alpha$ and consists of three compact knots A, B and C (knot C is very close to a  star)  connected with a nebulous filament.  HH~1117 and HH~1118 are located close to the axis of the HH~1115 jet and may trace additional shocks in the same outflow. HH~1119 consists of a pair of nebulous patches with a dumbbell shape brighter in [SII] than H$\alpha$ in a low extinction environment well  separated from the other HH objects around GM~3-7.

\begin{figure}
\includegraphics[width=80mm]{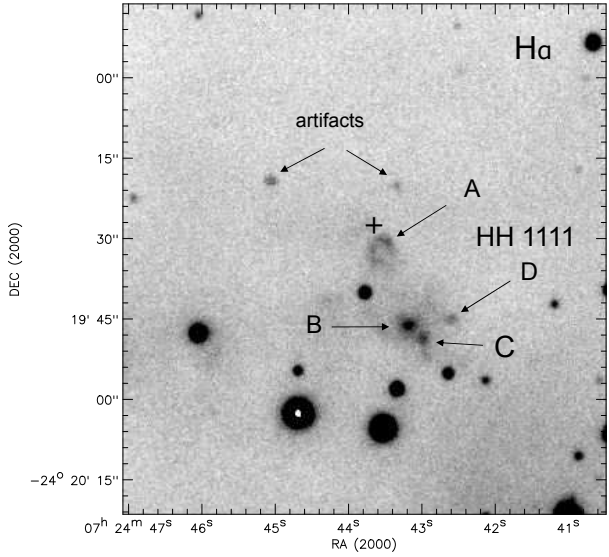}
\caption{The HH flow near embedded IR source in H$\alpha$ emission. The individual knots are marked by letters. The position of IR star is shown by cross.}
\label{HHflow}
\end{figure} 

\begin{figure}
\includegraphics[width=80mm]{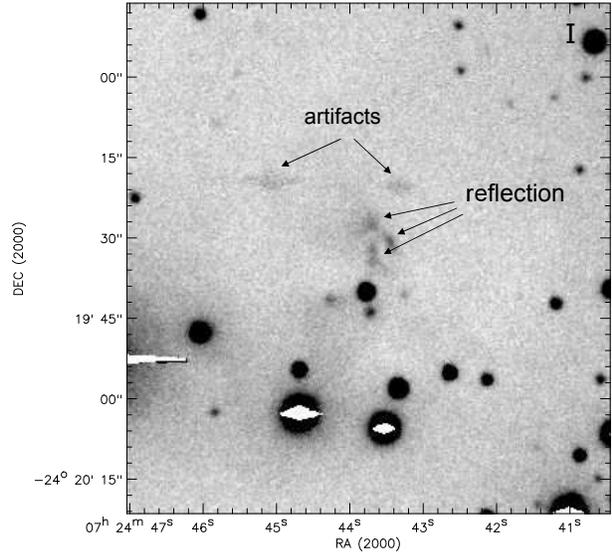}
\caption{The same area in the i' filter. Instead of HH knots, one can see reflection patches, tracing the conical cavity.}
\label{HHflow_i}
\end{figure} 
   
\subsubsection{A group near an embedded IR source}

HH~1111,  located near the edge of dark cloud far from the main area of HH activity in the L1664, is shown in Figure~\ref{HHflow}.
The  central star is  invisible at visual wavelengths, seen in the 2MASS H-band,   and especially bright in the K$_{s}$ image.    In [SII] and  H$\alpha$ several distinct emission knots are surrounded by extended nebulosity southwest of  the infrared star. The knots, designated  HH~1111 A, B, C, D in Figure~\ref{HHflow}, are totally absent in the i' image (Figure~\ref{HHflow_i}),   but a few faint patches of reflection nebulosity tracing the walls of a conical cavity created by the outflow can be seen near the position of the knot A.    The knots are distributed over a wide range of  angles ($\sim$45\degr ) with respect to the suspected source star.     Thus, HH~1111 seems to trace a poorly-collimated flow.  

\subsubsection{HH~1108}

\begin{figure}
\includegraphics[width=85mm]{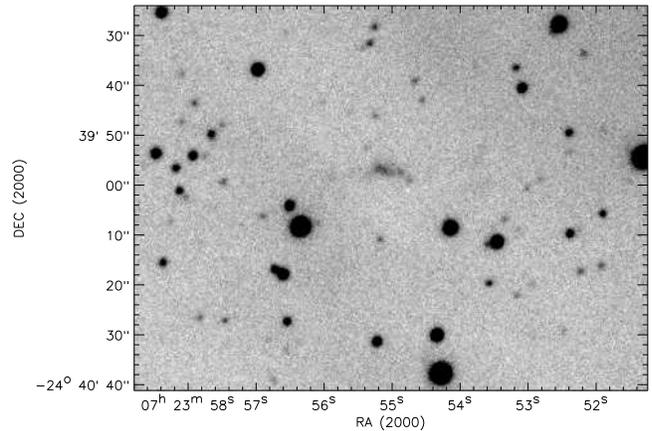}
\caption{The H$\alpha$ image of HH~1108.}
\label{HH1108}
\end{figure}

This object resembles a streak with several condensations projected onto a dark cloudlet. It is  brighter in H$\alpha$ than [SII] but  has similar  structure in both filters.   Nothing is detected in  i' or the 2MASS images.   HH~1108, shown in Figure~\ref{HH1108},  is located far from the areas described above.  

\begin{figure*}
\includegraphics[width=160mm]{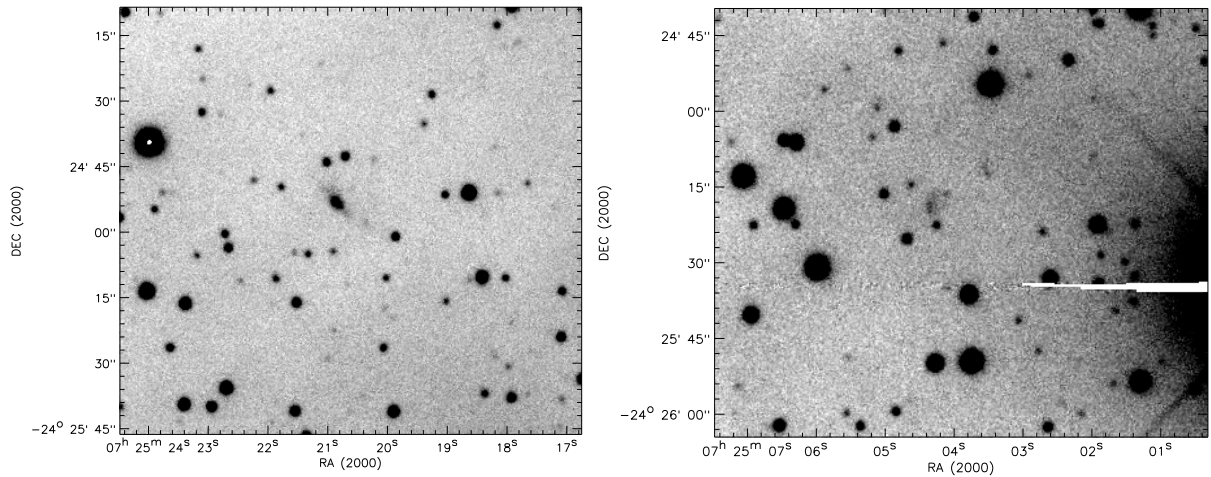}
\caption{The H$\alpha$ images of Em.Obj.~1(right panel) and Em.Obj.~2 (left panel).}
\label{EmObjs}
\end{figure*}

\subsubsection{Emission objects of unknown nature}

Two more  emission-line objects  shown  in Figure~\ref{EmObjs} and listed in the Table 1 were found that cannot be unambiguously classified as HH objects.   They might be planetary nebulae,   distant HII regions,   or  genuine HH objects, but without spectra,  classification is impossible.

\textit{Em. Obj. 1:  }This object is  a tight group of emission knots brighter in [SII] than H$\alpha$.    Nothing is  found in 2MASS or WISE at its location.   Although its remote  location from the cloud makes it unlikely to be an  HH object, it might be a remote, parsec-scale extension of  the flow originating from GM 2-21 (see above).   This object is located 300 arc-seconds from the apex of the cometary nebula GM 2-21 and within a few arc-seconds of the axis of the jet emerging from the embedded star.  

\textit{Em. Obj. 2: }  This object's appearance is  similar to an edge-on spiral galaxy, but it is invisible  in i'  and 2MASS.   Its emission line character suggests that it might be a distant planetary nebula. One should note that the  large diffuse object, seen in Fig.\ref{field} in 3 arc-minutes below the location of Figure 10a, and visible in both the narrow-band and the i' filters, is a background galaxy.

\begin{table*}
\label{RN_table}
 \centering
 \begin{minipage}{140mm}
  \caption{The coordinates of the reflection and cometary nebulae in the field.}
  \begin{tabular}{@{}lllll@{}}
  \hline
   Name     &  RA(2000)    & Decl.(2000) & Other names & Notes  \\
 \hline
 RN~5a     & 7 23 40.1 & $-$24 43 42 &  &  \multirow{2}{*}{$\Bigr\}$ Nebulous patch on the edge of dark cloudlet, with two stars} \\ 
 RN~5b     & 7 23 41.7 & $-$24 43 40 &  &   \\
 BBWo~22A & 7 23 51.9 & $-$24 46 00 &  & Three stars of nearly equal brightness surrounded by nebula \\
 BBWo~22B & 7 24 01.0 & $-$24 24 04 &  & Bright star with a very faint nebula \\
 RN~1     & 7 24 05.9 & $-$24 38 53 &  & Nebulous star near BBWo~22C \\
 BBWo~22C & 7 24 14.1 & $-$24 38 16 & GN~07.22.1, RK~71 & Includes IR cluster DBS~11 and IRAS~07221-2531 \\
 RN~2     & 7 24 22.5 & $-$24 35 49 &  & Nebulous star near BBWo~22C \\
 BBWo~22D & 7 24 31.1 & $-$24 41 37 &  & Bright star with a very faint nebula\\
 GM~1-8 & 7 24 37.0 & $-$24 34 47 & GLMP~181, PDS~250  & Associated with IRAS~07225-2428\\
 RN~6  & 7 24 38.8 & $-$24 28 11 & & Typical cometary nebula on NW from GM~3-7 \\ 
 GM~2-20 & 7 24 40.5 & $-$24 34 23 & BBWo~22E    \footnote{in SIMBAD is identified incorrectly}\\
 RN~3 & 7 24 40.8 & $-$24 35 33 &  & Filamentary reflection nebula \\
 RN~7 & 7 24 43.9 & $-$24 31 57 &  & A nebulous star \\
 RN~4 & 7 24 44.4 & $-$24 35 08 &  & A nebulous star \\
 RN~8 & 7 24 46.2 & $-$24 31 55 &  & A star with nebulous fan \\
 GM 3-7 & 7 24 46.7 & $-$24 30 00 & BBWo 22F \footnote{in SIMBAD is incorrectly identified with GM~2-21}  & Contains several stars  \\
 BBWo 22G & 7 24 47.5 & $-$24 41 20 &  & Bright star with a very faint nebula \\
 GM~2-21 & 7 24 51.4 & $-$24 29 23 & & Typical cometary nebula, a host for HH-jet HH~1115 \\
GM 1-22 & 7 24 54.0 & $-$24 29 30 & BBWo~22H  &  \\
\hline
\end{tabular}
\end{minipage}
\end{table*}

\subsection{Reflection and cometary nebulae}

The L1664 cloud contains many small reflection nebulae associated with one or more stars with strong infrared excess emission.  Thus, this cloud is an example of an association of  reflection nebulae (R-association in the older literature) tracing a loose group  low- to intermediate mass young stellar objects.   Some of these nebulae are associated with bright stars  listed in BBWo and GM catalogues but never studied in detail.    The new reflection nebulae reported here tend to trace objects whose illuminating stars are faint.    Some are associated with HH objects or located near other outflows described above.    All new and previously catalogued reflection nebulae are listed  in Table 2 with updated coordinates.
Short descriptions are given below.

\textit{RN~5: }
This nebula represents the edge of a dark cloudlet  illuminated by one or two stars.    RN~5  is located near the axis of the elongated ellipse which includes nearly all objects  in this field. No  emission lines  were detected.

\begin{figure}
\includegraphics[width=80mm]{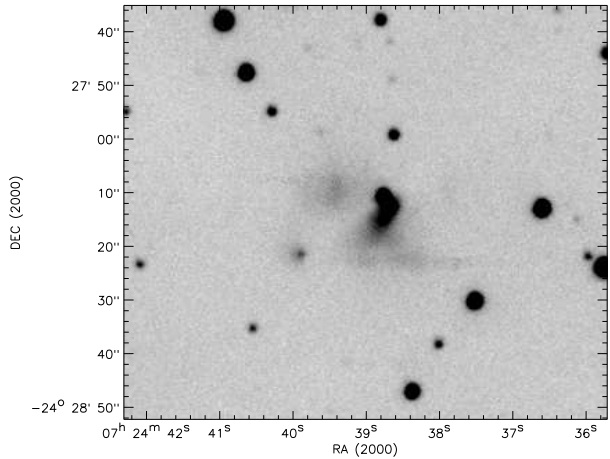}
\caption{RN~6 reflection nebula in i' filter.}
\label{RN6}
\end{figure}

\textit{BBWo~22B:}
This bright, slightly nebulous star is located far from the field containing the HH objects.   

\textit{RN~6:}
A compact and bright cometary reflection nebula with  traces of bipolar structure.
Its central star is located near the edge of a region  of  high extinction (the same region includes GM~3-7) but it is  not noticeably reddened (Figure~\ref{RN6}).

\textit{RN~7 and RN~8:}
These are two nebulous stars, detected to the south of HH~1114.

\textit{BBWo~22G:}
This rather bright star with wisp of nebulosity, similarly to BBWo~22B, is aside from the main area of HH-activity.

\begin{figure*} 
\includegraphics[width=180mm]{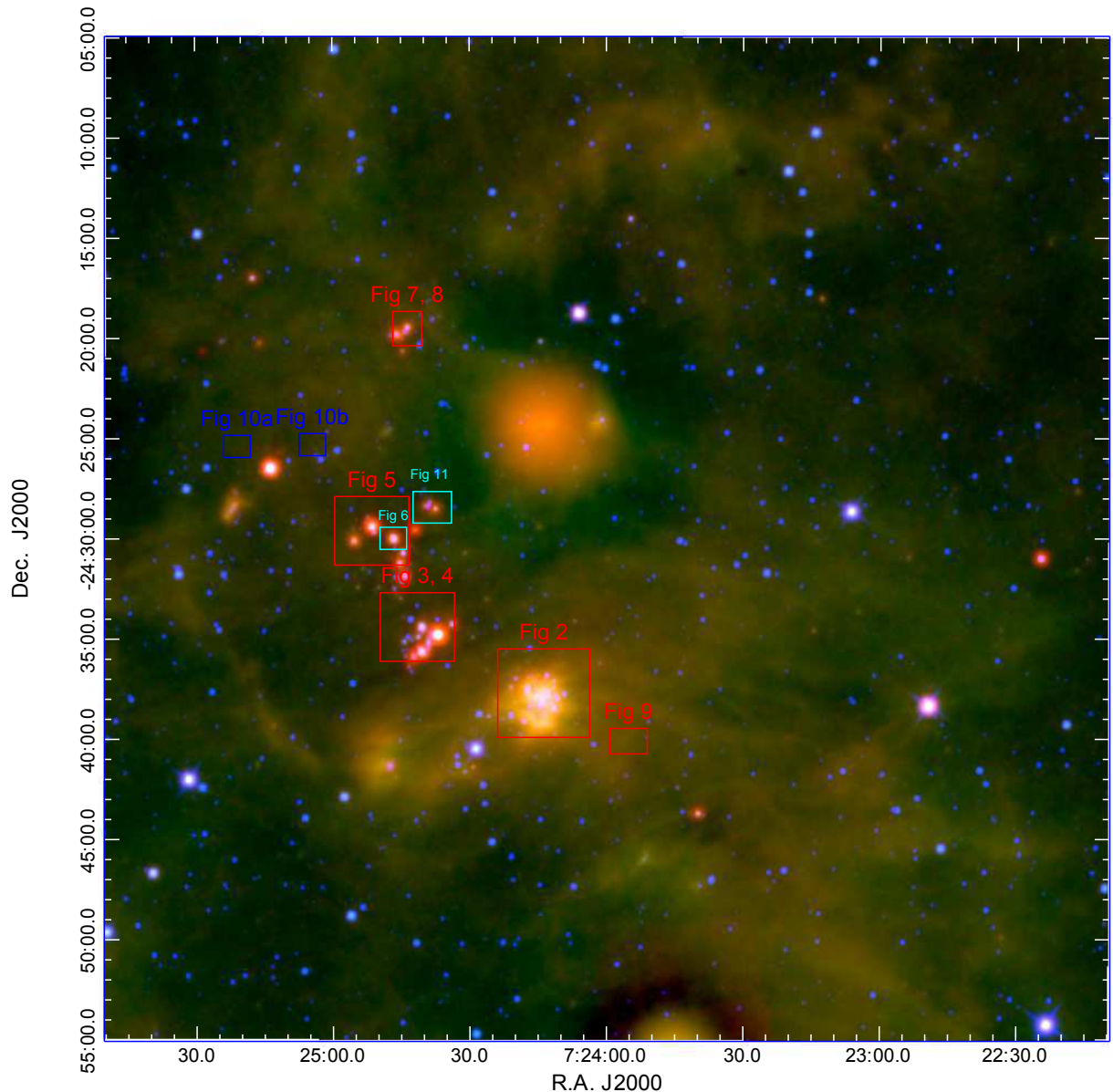}
\caption{A WISE Band 2 (4.6 $\mu$m; blue), Band 3 (12 $\mu$m;  green), and Band 4 (22 $\mu$m; red) image showing the field around Lynds 1664.   As in Figure 1, the boxes mark indicate the locations of the various figures.}
\label{WISE}
\end{figure*}

Figure \ref{WISE} shows a thermal-infrared image obtained by the WISE satellite at 4.6 $\mu$m,  12 $\mu$m, and 22 $\mu$m.   This image shows warm dust associated with BBWo 22C (Figure 2) ,  a cluster of more than a dozen YSOs in the field containing GM 1-8 and GM 2-20 (Figures 3 and 4), about a dozen YSOs associated with GM 1-22, GM 2-21, and GM 3-7 (Figure 5), and and small group of YSOs in the field containing HH~1111 (Figures 7 and 8).

\section{Discussion and Conclusions}

The limited available observations do not allow a thorough analysis of this star-forming region whose distance remains unknown.   The small sizes of the reflection nebulae and  HH objects, the high density of foreground stars despite the anti-center direction location of L1664 suggest that it is much  farther than the  Perseus, Orion,  or  the North America / Pelican Nebula complex in Cygnus.  Several other dark clouds such as  L1667, which contains the GM 1-46 nebula \citep{GM1} and the Bok globule L~1660, which contains HH~72 and several emission-line young stars and reflection nebulae \citep{reipurthgraham} are located within a half degree of the L1664 dark cloud.    The [SII] and especially the H$\alpha$ images show extended diffuse emission on the Western side of L1664.   This emission may belong to the large and diffuse HII region Sh-2 310.     It is difficult to determine  where the L1664 dark clouds  are located -- in foreground or behind the emission from Sh-2 310.   However, the distances to these clouds and the HII region Sh-2 310 are controversial.     \citet{reipurthgraham}  assumed that they were related to the Vela~OB1 association at a distance of  1.5 kpc.       \citet{pds1} studied the central star (PDS~250) in GM~1-8 and its distance was  found to be  2.46 kpc (kinematic distance from radio observations) or  even 4.1 kpc (photometric distance).

L1664 dark cloud is associated with at least two photometrically studied IR star clusters.    DBS~11,  located inside the BBWo~22C reflection nebula and connected to two HH objects (see above), is definitely associated with  L1664.  Its (J$-$H),(H$-$K$_s$) diagram \citep{soares2} demonstrates that the cluster  contains  low-mass young stars.  Unfortunately, the  distance was not estimated.    The range of A$_V$ from 9 to 17 magnitude, given  by \citet{ivanov}  seems overestimated given their visibility in the images presented here.  

\citet{soares1} studied the small cluster surrounding the GM~2-20 and GM~1-8 nebulae  (see Figure~\ref{GM2-20}).   Using the  J vs. (J$-$H) color-magnitude diagram,  they estimate the distance modulus J$-$M$_J$ as  10.5 implying a distance of 1.3 kpc.     Though this value is  lower than estimated by  \citet{pds1}, it seems in good agreement with the distance of Vela~OB1.    Finally, the distance to the well-studied cluster NGC~2362,  believed to ionize the Sh-2~310 HII region, is established as 1.5 kpc \citep{dahm}.

\citet{kumar} investigated the physical sizes of embedded young clusters, finding that half of the sample have effective radii of 1.5 pc.    The half-light radii of the three small clusters in L1664 are about 100 arc-seconds.  However, given the small numbers of stars, this is a highly uncertain number.  If this corresponds to a length-scale of 1.5 pc, then the implied distance to L1664 is 2.1 kpc.

Thus, we assume that L~1664 and all objects inside it are located at D = 1.5 kpc.    The full projected length of the long axis of the star forming region is about 8 pc.   The longest   confirmed flow in the field -- from GM~2-21 to HH~1118 --  has a projected length of 0.75 pc at this distance.  If the HH object candidate Em.Obj.2 is confirmed to be an HH object and traces a distant shock in the outflow from GM~2-21, then this lobe of the outflow has a projected length of about   2.2 pc.  If the cloud is at a distance of 2.1 kpc, all dimensions are lager by a factor of 1.4.

\section*{Acknowledgments}

We thank Prof. Bo Reipurth for providing the numbers for new HH objects and  for some helpful suggestions.

This publication makes use of data products from the Two Micron All Sky Survey, which is a joint project of the University of Massachusetts and the Infrared Processing and Analysis Center/California Institute of Technology, funded by the National Aeronautics and Space Administration and the National Science Foundation.

This publication makes use of data products from the Wide-field Infrared Survey Explorer, which is a joint project of the University of California, Los Angeles, and the Jet Propulsion Laboratory/California Institute of Technology, funded by the National Aeronautics and Space Administration.

This work was partly supported by a research grant from the Armenian
National Science and Education Fund (ANSEF), based in New York, USA.

\label{lastpage}

\end{document}